\documentclass[12pt,twoside]{article}
\usepackage{fleqn,espcrc1,epsfig}

\usepackage{graphicx}
\usepackage[figuresright]{rotating}
                                                                           
\def\Journal#1#2#3#4{{#1} {\bf #2}, #3 (#4)}

\def\NIM{\em Nucl. Instrum. Methods}
\def\NIMA{{\em Nucl. Instrum. Methods} A}
\def\NPA{{\em Nucl. Phys.} A}
\def\NPB{{\em Nucl. Phys.} B}
\def\PLB{{\em Phys. Lett.}  B}
\def\PRL{\em Phys. Rev. Lett.}
\def\PRC{{\em Phys. Rev.} C}

\def\ZPA{{\em Z. Phys.} A}
\def\EPJA{{\em Eur. Phys. J.} A}

\newcommand{\peep}{$p(\vec e,e'\vec p\,)$~} 
\newcommand{\kle}{\raisebox{-2.5pt}{$^{<}$}}
\begin{document}

\title{The focal plane proton-polarimeter for the 3-spectrometer setup at
       MAMI}

\author{
Th.~Pospischil\address{Institut f\"ur Kernphysik, Universit\"at Mainz, 55099 Mainz, Germany}\footnote{comprises part of doctoral thesis}, 
P.~Bartsch$^{\rm a}$, D.~Baumann$^{\rm a}$, R.~B\"ohm$^{\rm a}$, 
K.~Bohinc$^{\rm a,}$\address{Institut Jo\v zef Stefan, University of Ljubljana, SI-1001 Ljubljana, Slovenia}, 
N.~Clawiter$^{\rm a}$,
M.~Ding$^{\rm a}$, S.~Derber$^{\rm a}$, M.~Distler$^{\rm a}$, 
D.~Elsner$^{\rm a}$, 
I.~Ewald$^{\rm a}$, 
J.~Friedrich$^{\rm a}$, 
J.M.~Friedrich$^{\rm a}$\footnote{present address: Physik Department E18, TU M\"unchen, Germany}, 
R.~Geiges$^{\rm a}$,
S.~Gr\"ozinger$^{\rm a}$, 
M.~Hamdorf$^{\rm a}$, 
S.~Hedicke$^{\rm a}$,
P.~Jennewein$^{\rm a}$, 
J.~Jourdan\address{Dept. f\"ur Physik und Astronomie, Universit\"at Basel, CH-4056 Basel, Switzerland}, 
M.~Kahrau$^{\rm a}$,
F.~Klein$^{\rm a}$, 
P.~K\"onig$^{\rm a}$, 
A.~Kozlov$^{\rm a}$\footnote{on leave from: School of Physics, University of Melbourne, Victoria, Australia},
H.~Kramer$^{\rm a}$,
K.W.~Krygier$^{\rm a}$, 
J.Lac\address{Rutgers University, Piscataway, NJ, USA}, 
A.~Liesenfeld$^{\rm a}$, S.~Malov$^{\rm d}$, J.~McIntyre$^{\rm d}$, 
H.~Merkel$^{\rm a}$, P.~Merle$^{\rm a}$,
U.~M\"uller$^{\rm a}$, R.~Neuhausen$^{\rm a}$, 
E.A.J.M.~Offermann$^{\rm a}$,
M.~Potokar$^{\rm b}$, R.~Ransome$^{\rm d}$, 
A.~Richter$^{\rm a}$,
G.~Rosner$^{\rm a}$\footnote{present address: Dept. of Physics \& Astronomy, 
                       University of Glasgow, UK }, 
J.~Sanner$^{\rm a}$, 
H.~Schmieden$^{\rm a}$\footnote{corresponding author (email: hs@kph.uni-mainz.de)}, 
M.~Seimetz$^{\rm a}$, 
I.~Sick$^{\rm c}$, 
O.~Str\"ahle$^{\rm a}$,
A.~S\"ule$^{\rm a}$, 
A.~Wagner$^{\rm a}$, Th.~Walcher$^{\rm a}$, 
G.A.~Warren$^{\rm c}$,
M. Weis$^{\rm a}$, X.-Q.~Wu$^{\rm d}$
        }


\maketitle

\begin{abstract}
For experiments of the type $A(\vec e,e'\vec p\,)$  
the 3-spectrometer setup of the A1 collaboration at MAMI has been 
supplemented by a focal plane proton-polarimeter.
To this end, a carbon analyzer of variable thickness
and two double-planes of horizontal drift chambers
have been added to the standard detector system of Spectrometer A. 
Due to the spin precession in the spectrometer magnets, 
all three polarization components at the
target can be measured simultaneously. 
The performance of the polarimeter has been studied 
using elastic \peep scattering.

~
keywords:\,\, proton polarimeter, drift chamber, spin precession,
              analyzing power
\end{abstract}

\section{Introduction}

At the high luminosity and high duty factor electron accelerators it has 
become possible in the last years to fully exploit the potential of 
recoil polarimetry in electron scattering.
This has led to interesting new results concerning the nucleon's
ground state and resonance structure.
Quasielastic scattering $D(\vec e, e'\vec n)$  experiments 
have proven the neutron electric form factor
to be substantially larger than previously assumed from unpolarized
measurements \cite{Ostrick99,Herberg99,HS99-Bates25,Eden94}.
At high momentum transfers, the measurement of recoil polarization in 
elastic electron-proton scattering has confirmed with high accuracy
that the proton electric form factor is approximately a factor
of two below the scaled magnetic form factor \cite{Jones00}.
Reaction mechanism and nuclear structure effects have been investigated 
in the $D(\vec e, e'\vec p\,)$  \cite{Eyl95,Milbrath98,KOSW98,Barkhuff99},
$^4\mbox{He}(\vec e, e'\vec p\,)$  \cite{Ransome99},
$^{12}\mbox{C}(\vec e, e'\vec p\,)$  \cite{Woo98} and
$^{16}\mbox{O}(\vec e, e'\vec p\,)$  \cite{Malov00} experiments.

In the $N$ to $\Delta$  transition, which is tagged through the 
$p(e, e'p)\pi^0$  reaction, 
recoil polarization has been shown to be sensitive to the small 
longitudinal quadrupole mixing \cite{Lourie90,HS98}.
While an experiment at MIT-Bates was only performed with unpolarized electron 
beam \cite{Warren98}, a polarized beam program is underway at
the Thomas Jefferson National Accelerator Facility (TJNAF)
\cite{JLab-N-Delta}, and first results are available from the Mainz
microtron MAMI \cite{HS00-Panic99,HS00-NSTAR2000,Pospischil00,Pospischil00a}.
In the parallel kinematics of the MAMI experiment the ratio $R_L / R_T$  
of longitudinal to transverse response can be 
furthermore extracted from the simultaneously
measured recoil polarization components without the need of a 
Rosenbluth-separation \cite{HS00-polsum}.

This paper reports on the proton polarimeter which was built for the 
3-spectrometer setup \cite{Blomqvist98} of the A1-collaboration at MAMI.
It is organized as follows:
Section \ref{sec:Method_Setup} describes the method of polarization 
measurements and the setup we chose for proton polarimetry
behind the focal plane of one of our spectrometers.
The horizontal drift chambers (HDCs) of the polarimeter are introduced in 
detail in paragraph \ref{sec:HDC}. 
Section \ref{sec:polarization} describes calibration measurements of 
the proton polarization in the elastic $p(\vec e, e'\vec p\,)$  reaction.
The spin precession in the spectrometer, instrumental asymmetries and the
absolute calibration are discussed. 
A short summary finally is given in section \ref{sec:Summary}.

\section{Method and setup}
\label{sec:Method_Setup}

In all the above mentioned experiments the polarization of the recoiling
nucleons is measured through secondary scattering in a strong-interaction
process.
The strong spin-orbit coupling causes an azimuthal asymmetry from which
the polarization perpendicular to the nucleon momentum can be extracted.

Polarimetry is often performed after a
momentum-analyzing magnetic deflection of the protons in a spectrometer
\cite{Bratashevsky80,Kato-Takeda80,McClelland84,Haeusser87,Eganov96}.
This also automatically provides the spin-precession which enables the
measurement of the longitudinal polarization component.
At the same time it causes a mixing of the polarization components which needs
to be disentangled later on.

Except for liquid helium
at high proton energies \cite{Kato-Takeda80},
the focal plane proton polarimeters usually use carbon as analyzer, 
because it is easy to handle and 
the inclusive scattering of polarized protons on carbon has an analyzing 
power $A(\Theta_s,T_p)$  
which is experimentally well known as a function of the proton
kinetic energy, $T_p$, and scattering angle, $\Theta_s$
\cite{Aprile-Giboni83,McNaughton85}.
From the modulation of the $^{12}$C$(\vec p,p')$  cross section 
with the azimuthal angle, $\Phi_s$,  around the polarization independent
part, $\sigma_0(\Theta_s,T_p)$,
\begin{equation}
\label{eq:sigma}
\sigma = \sigma_0(\Theta_s,T_p)\left[ 1 + A(\Theta_s,T_p) \left( P_y^{fp} \cos{\Phi_s} - P_x^{fp} \sin{\Phi_s} \right) \right],
\end{equation}
it is possible to extract two polarization components $P_x^{fp}$  and
$P_y^{fp}$, which in the focal plane are oriented
perpendicular to the proton momentum.
The reconstruction of the polar and azimuthal scattering angles
requires proton tracking before and after scattering.
Thus, for recoil proton polarimetry in electron scattering coincidence 
experiments, 
spectrometers have been equipped with polarimeters made up of a carbon analyzer
sandwiched by tracking detectors \cite{Woo98,CEBAF-Pol}.

\begin{figure}[t]
\centerline{\psfig{figure=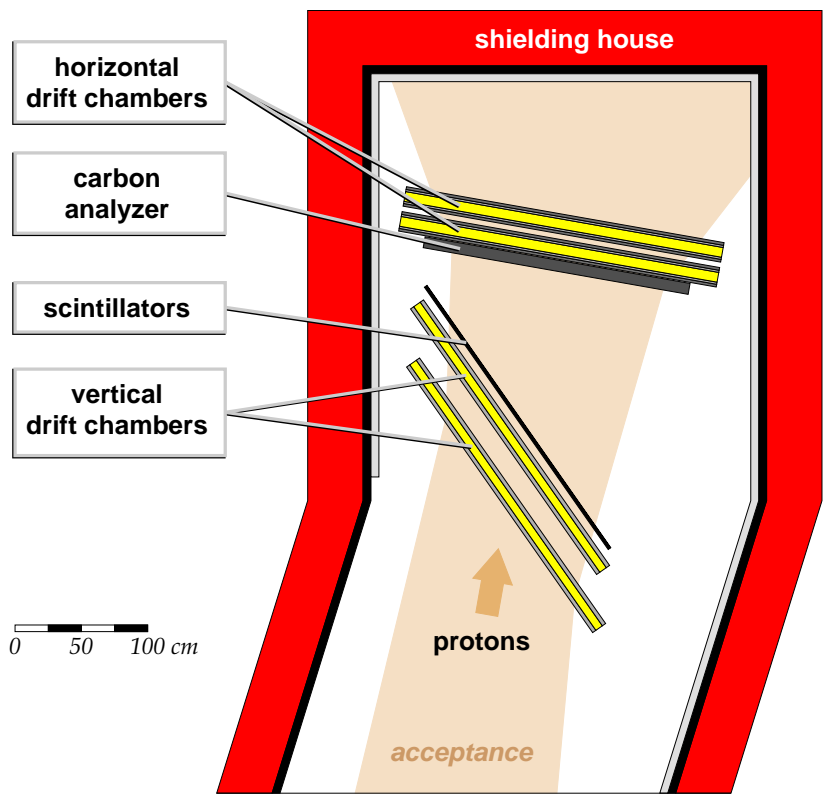,width=12cm}}
\caption{ Setup of the proton polarimeter in the shielding house of
          Spectrometer A. The standard detector system is supplemented 
          by a carbon analyzer and two double-planes of HDCs for proton
          tracking after scattering. }
\label{fig:setup}
\end{figure}

In the case of the 3-spectrometer setup at MAMI the standard
focal plane detectors of Spectrometer A consist of two double-planes
of vertical drift chambers (VDCs) and two 3\,mm and 10\,mm thick layers
of plastic scintillators for timing purposes and particle identification
\cite{Blomqvist98}. 
These detectors are also used for proton tracking before scattering 
from carbon.
They are supplemented by the carbon analyzer followed by two double-planes 
of horizontal drift chambers (HDCs) as is illustrated in 
Figure\,\ref{fig:setup}.
\input{psfig}
The shielding house of Spectrometer A is indicated in dark grey. 
The light-grey shaded band indicates possible proton trajectories.
They cross, from bottom to top, the two VDCs and the scintillators, 
and then impinge on the graphite analyzer.
Its thickness can be optimized between 1 and 7\,cm 
(density $\rho = 1.76$\,g/cm$^3$)
for protons up to the 
spectrometer's maximum central momentum of 660\,MeV/c.
With an active area of $2178 \times 749.5$\,mm$^2$  the HDCs are 
large enough to measure proton scattering angles of up to $20^{\circ}$  
over the full size of the carbon analyzer.
This covers the region of high analyzing power.
Even between $20^{\circ}$  and $35^{\circ}$, $97.7\,\%$  of the scattered
protons are geometrically accepted.
The HDCs are the crucial new parts of the polarimeter setup.
They are described in detail in the next section.

\section{Horizontal drift-chambers}
\label{sec:HDC}

The polarimeter HDCs realize a simple geometry which is similar to
early designs \cite{Walenta71,Sauli77}.
The electric field is formed by alternating so-called potential wires
and signal wires.
In our case the former are grounded whereas the latter carry positive high
voltage of typically 3000\,V.

All wires of the polarimeter HDCs
are gold-plated tungsten with diameters of 50 and 100\,$\mu$m
for the signal and potential wires, respectively.
Each wire plane consists of 103 signal wires and 104 potential wires.
Their maximum length is 106\,cm because they are stretched under 45$^\circ$
across the wire frames.
The wires of the two individual planes of a double-plane are perpendicular to 
each other.
\begin{figure}[t]
\centerline{\psfig{figure=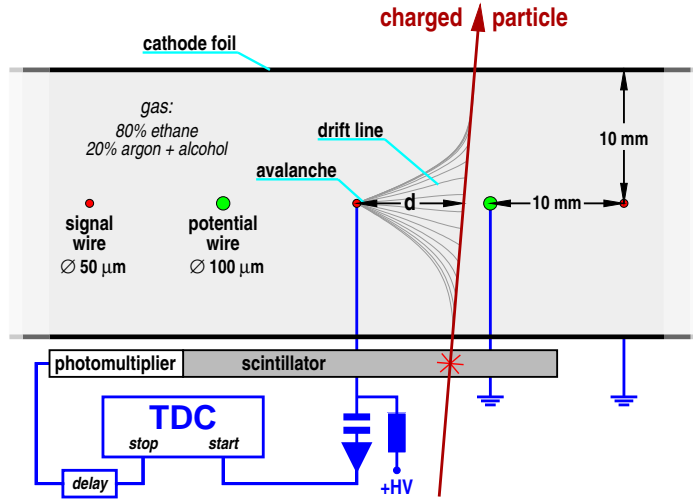,width=10cm}}
\caption{ Schematics of the HDC. The field forming signal and potential 
          wires have a distance of 1\,cm to each other and to the
          cathode foils. Electrons from the ionization along the particle
          track drift to the signal wire. The drift time
          measurement is started by the individual signal wires and 
          stopped by a fast external trigger scintillator.}
\label{fig:HDC_schematics}
\end{figure}
As can be seen from the schematical drawing of Figure\,\ref{fig:HDC_schematics}
the wire separation is 10\,mm.
This is also the distance between the wire plane and the cathode foils, 
which consist of 6\,$\mu$m {\it Mylar}\footnote{registered trademark of DuPont} 
with double sided aluminium coating.
A single drift cell has a cross section of $20 \times 20$\,mm$^2$.

An incoming charged particle creates electron-ion pairs along its track.
With our gas composition of 20\,\% argon\footnote{saturated with ethanol
at room temperature} 
and 80\,\% ethane protons of 150\,MeV kinetic energy have a 
specific energy loss of $dE/dx = 7.6$\,keV/cm at normal pressure.
This results in approximately 290 electron-ion pairs per cm. 
The electrons drift to the nearest signal wire, in the vicinity of which 
the gas amplification occurs.
The signals are fed through a high voltage capacitor to standard
{\it LeCroy}\,2735DC amplifiers/discriminators, whose outputs then start
time-to-digital converters (TDC) of a TDC2001 system \cite{Clawiter95}
individually for each signal wire.
The drift time is measured against the standard trigger-scintillator plane
of Spectrometer A which stops the TDC after an appropriate delay.

\begin{figure}[t]
\begin{center}
\epsfig{file=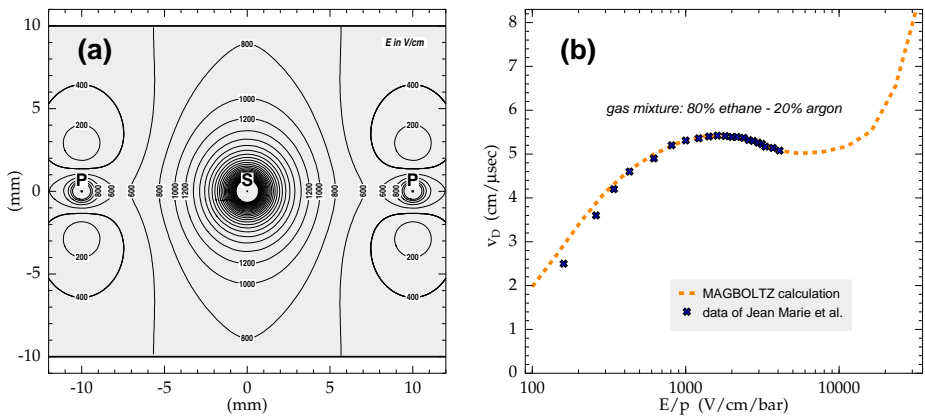, width=12.5cm} 
\caption{ (a) Contour lines of the electric field strength in a drift cell.
          Cathode foils and potential wires (P) grounded, signal wires (S)
          on +3000\,V. Calculated with the computer code \texttt{GARFIELD}
          \cite{Veenhof99}.
          (b) Electron drift velocity in a gas mixture of
          80\,\% ethane and 20\,\% argon as a
          function of pressure-normalized electrical field strength.
          Calculation (dashed line) with the computer code \texttt{MAGBOLTZ}
          \cite{Biagi97} compared to measurement (crosses) 
          \cite{JLH79}.   }
\label{fig:HV-v_drift}
\end{center}
\end{figure}

With known drift velocity of the electrons the TDC information can be 
converted into drift distance.
For the given gas composition the drift velocity, $v_D$, depends on
the reduced field strength, $E/p$.
Both the field strength in the drift cell and the drift velocity 
are shown in Figure\,\ref{fig:HV-v_drift}.
Except at the `corners' above and below the potential wires the field strength 
is in the range 0.4 -- 15\,kV/cm.
Thus, at normal pressure the drift velocity has values within  
$\pm 10$\,\% around $v_D \simeq 5$\,cm/$\mu$s;
this plateau makes the operation of the HDC
insensitive against small changes of the external conditions, e.g.
air pressure and high voltage.

At a field strength of 1\,kV/cm 
the longitudinal diffusion broadening 
is only 100\,$\mu$m per cm of drift \cite{JLH79}, which is a factor of ten
better than in pure argon.
Furthermore,  with 80\,\% ethane as photon quencher
a gas amplification of $10^4$  -- $10^6$  
can be achieved with well localized avalanches, 
which is 1 -- 2 orders of magnitude higher than in pure argon.
The localization of the avalanches plays a crucial role for the left-right
assignment in the HDC.

\subsection{Left-right assignment}

A standard problem in HDCs is the left-right ambiguity:
From the measurement of a single drift time it cannot be decided whether the 
particle track occured left or right of the signal wire.
However, if the avalanches are well localized on the particle track's 
side of the signal wire \cite{FOW78}, 
different signals are induced on the two potential wires bounding a 
drift cell;
the signal is larger on that side of the signal wire where the avalanche 
occured \cite{Walenta78}.
The potential wire signals are approximately an order of magnitude below those
of the signal wires, and the difference between the potential wire signals 
is another factor of ten smaller.
Assuming a few hundred avalanches of about 
$5 \times 10^5$  electron-ion pairs from a particle track,
for the given geometry and operating conditions 
a difference signal of $\Delta I \simeq 100$\,nA is obtained
over a time-interval of 200 -- 300\,ns. 

\begin{figure}[t]
\centerline{\psfig{figure=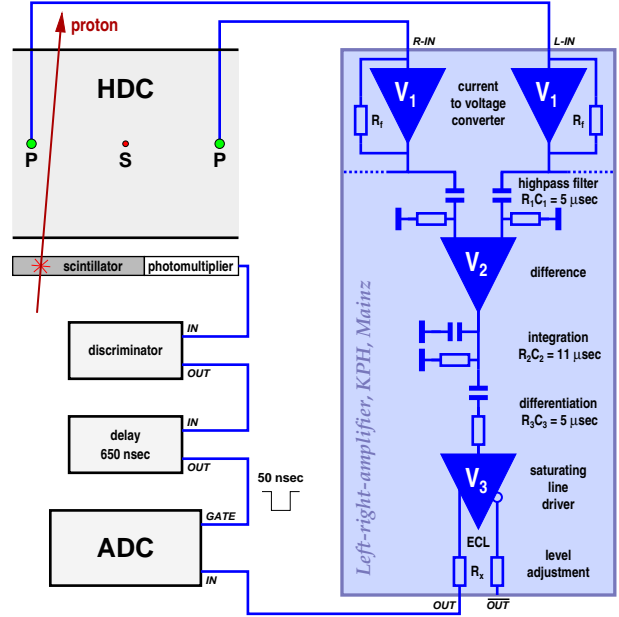,width=8.5cm}}
\caption{ The left-right amplifier's principle of operation. Signals of 
          adjacent potential wires after linear amplification are fed to a
          differential amplifier. The output signal is integrated in an ADC. }
\label{fig:amplifier}
\end{figure}

In order to exploit
the small difference between the potential wire signals 
a special so-called left-right amplifier has been designed and built 
\cite{Jennewein95,JB_94-95}.
Its circuit diagram is schematically depicted in Figure\,\ref{fig:amplifier}.
The currents induced on adjacent potential wires first are converted to a 
voltage by the amplifiers V$_1$.
The feedback resistors R$_f$  have to be equal within $0.1\,\%$  
in order to achieve a common mode rejection of typically 50\,dB 
for the differential amplifier V$_2$.
Due to the low input-impedance of V$_1$  a `sectoring',
i.e. the combination of the potential wires from several cells,
is possible:
Each wire plane is divided into 10 odd-even sectors of 7 (full-length)
up to 20 (shorter - in the corners of the HDC) drift cells 
in which all odd and even numbered potential wires, respectively, 
are bussed together into one input of V$_1$.
An additional left-right amplifier is used for each drift cell
inbetween two sectors. 
Therefore the whole plane is read out by 19 `odd-even' amplifiers.

The output signal of the differential amplifier V$_2$  in 
Figure\,\ref{fig:amplifier} is integrated and then differentiated 
in order to achieve a recovery time of better than 
20\,$\mu$s after overload.
After further amplification and level shifting in V$_3$, the output pulses
are fed to a 96-channel {\it LeCroy} 1882N analog-to-digital converter (ADC). 
There the signal is integrated for 50\,ns.
The ADC gate is delayed by 650\,ns
relative to the trigger scintillator.

\begin{figure}
\centerline{\psfig{figure=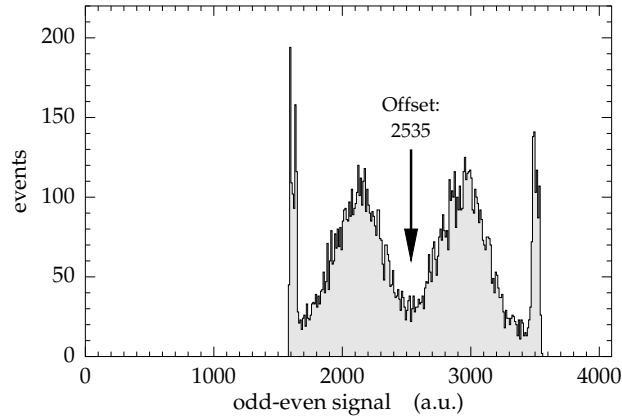,width=8.5cm}}
\caption{ ADC-spectrum for one odd-even sector. }
\label{fig:ADC_odd-even}
\end{figure}

A typical odd-even spectrum is shown in Figure\,\ref{fig:ADC_odd-even}.
If the current-difference between odd and even wires of a sector
$I_{odd} - I_{even} < - 90$\,nA, then the output of the odd-even amplifier 
is in negative saturation which shows up as the right peak in the ADC spectrum.
For $I_{odd} - I_{even} > 90$\,nA the amplifier is in positive saturation 
and the corresponding events are located in the left peak of 
Figure\,\ref{fig:ADC_odd-even}.
The amplifier's output is proportional to the input-current difference
when $|I_{odd} - I_{even}| < 90$\,nA.

Events in the left and right parts correspond to tracks through the `even' 
and `odd' side of the sector, respectively.
Entries around the central minimum of spectrum 
are mainly due to tracks close to the signal wire.
The good performance of the left-right decision was confirmed through 
measurements without carbon analyzer, 
both with the large HDCs and with a small prototype HDC \cite{Hamdorf96}.

\subsection{Drift time and drift distance}

\begin{figure}
\centerline{\psfig{figure=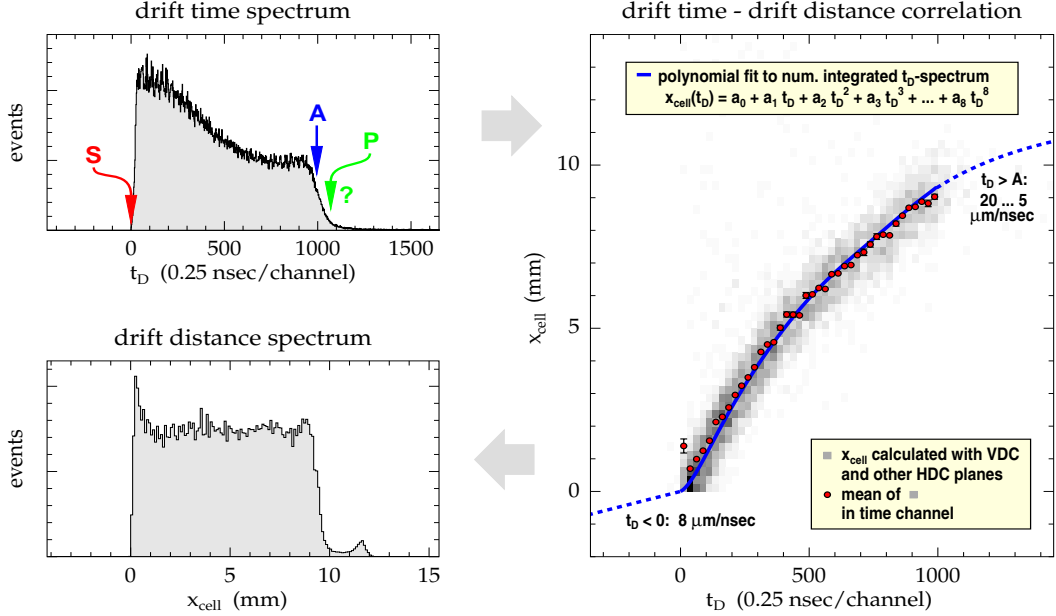,width=15cm}}
\caption{ Drift-time (top left) and drift-distance (bottom left) spectra
          and relation between time and distance (right) as obtained by 
          numerical integration of the time spectrum. 
          For explanation see text. }
\label{fig:time-distance}
\end{figure}

After the correct left-right assignment the drift time can be interpreted 
in terms of a position coordinate perpendicular to the actual wire direction.
Despite the plateau in the drift-velocity distribution of 
Figure\,\ref{fig:HV-v_drift} the assumption of a constant drift velocity is too
rough an approximation.
If the drift cell is uniformly illuminated, a detailed relation between 
drift time and drift distance can be established from the drift-time
spectrum (cf. Figure\,\ref{fig:time-distance} top left) itself.
The drift times cover a range of 0 -- 250\,ns which corresponds to the wire 
spacing of 1\,cm and the average drift velocity of almost 5\,cm/$\mu$s.
The non-flatness of this spectrum reflects the differential deviations of 
the drift velocity from the mean value.

Each drift-time interval $[t_D,t_D+\Delta t_D]$ can be attached to a 
drift-distance interval $[x_{cell},x_{cell}+\Delta x_{cell}]$.
The number of events in this time interval, $\Delta N(t_D)$,
is given by the number of events in the correponding drift-distance 
interval, $\Delta N(x_{cell})$, 
and the time interval is related to the drift-distance interval through
the local drift velocity,
$v_D(x_{cell})$:
\begin{equation}
\frac{\Delta N(t_D)}{\Delta t_D} = \frac{\Delta N(x_{cell})}
                                        {\Delta x_{cell} / v_D(x_{cell})}
       = \frac{\Delta N(x_{cell})}{\Delta x_{cell}} \cdot
         \frac{\Delta x_{cell}(t_D)}{\Delta t_D   }.
\end{equation}
Uniform illumination of the drift cells yields constant
$\Delta N(x_{cell}) / \Delta x_{cell}$.
Therefore the relation between $x_{cell}$  and $t_D$  can be determined by
integration of the drift-time spectrum:
\begin{equation}
x_{cell}(t_D) = x_{cell}(A) \cdot
                \frac{\int_0^{t_D} \frac{dN(t)}{dt} dt}
                     {\int_0^{t_D(A)} \frac{dN(t)}{dt} dt}.
\label{eq:x-cell}
\end{equation}
The lower limit of integration, $t_D = 0$, is related to tracks
directly at the signal wire (indicated by S in the drift-time spectrum
of Figure\,\ref{fig:time-distance}). 
However, the upper limit, $t_D(A)$, is not very well determined due to
the decrease of efficiency in the regions of reduced field
strength close to the potential wires (compare Figure\,\ref{fig:HV-v_drift}a).
Therefore the `position' of the potential wire in the drift-time spectrum
(indicated by P) is not well defined.
Instead, as the upper integration limit in Eq.\,\ref{eq:x-cell}
the edge indicated by A is used which corresponds to the decline of the 
efficiency.
\begin{figure}[t]
\centerline{\psfig{figure=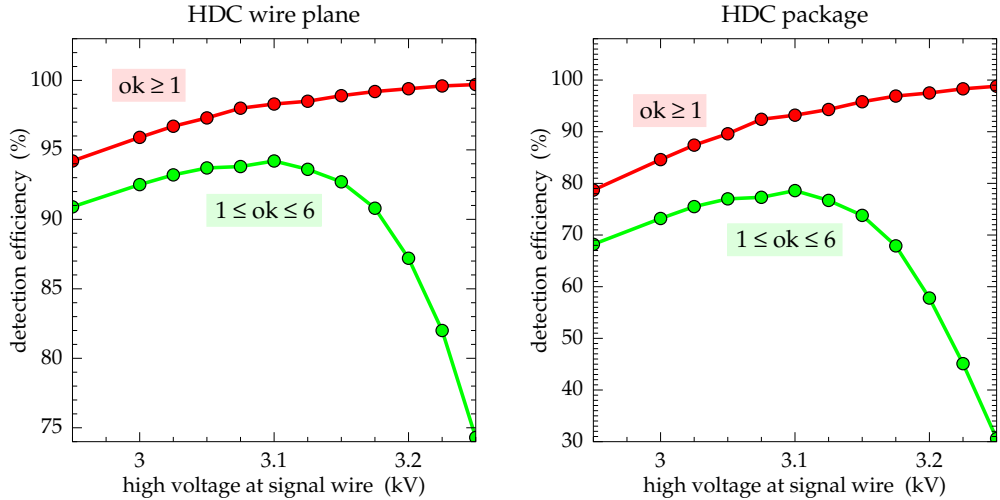,width=14cm}}
\caption{ HDC efficiency as a function of high voltage for 
          protons of $T_{HDC}=130$\,MeV kinetic energy 
          and a rate of $2.5$\,kHz at the HDC, and a
          discriminator threshold adjusted to $U_{thres.}=7$\,V at the
          {\it LeCroy}\,2735DC amplifier/discriminator (corresponding to
          a current threshold of 14\,$\mu$A).
          The left part is for one single plane, the projection to the 
          total efficiency of the 4-plane package is shown right.
          The $ok$  classification is described in the text.  }
\label{fig:efficiency}
\end{figure}

The drift-time to drift-distance relation is fitted by an 8$^{th}$  order
polynomial, which is shown as full curve in the right part of 
Figure\,\ref{fig:time-distance}.
It is confirmed by extrapolating particle trajectories measured with
the VDCs to the bottom plane of the HDC-package of the polarimeter.
These extrapolation results are indicated grey in the right plot,
and the column-wise mean values are represented by the points.
For standard operating conditions
best agreement between measurement and the numerical integration
in Eq.\,\ref{eq:x-cell} is obtained with 
$x_{cell}(A) = 9.3$\,mm. 
The continuation of the polynomial over the bounding of the drift cell 
in order to get a continuous relation is somewhat arbitrary.

From the drift-time to drift-distance relation the drift-distance 
spectrum (bottom left in Figure\,\ref{fig:time-distance}) is obtained.
Events at $x_{cell} > 11$\,mm are due to very long drift times 
$t_D > 400$\,ns.
They are attributed to tracks through the edges of the drift cells where
the field strength is low.
The peak at small distances between 200 and 500 $\mu$m is due to the fact
that for particle tracks with zero distance to the signal wire a sufficient
number of avalanches does not occur before a short delay.
This effect does not produce errors larger than the 
position resolution of the HDCs, which was determined to
$\delta x \simeq 300$\,$\mu$m from measurements of proton tracks 
with the prototype HDC relative to the standard VDCs.
With the distance of 22\,cm between the two HDC double-planes
an angular resolution of approximately 2\,mrad is achieved.

\subsection{HDC efficiency}
\label{sec:efficiency}

The efficiency of three of the four HDC planes can be measured by a
`sandwich' method:
If both the standard VDCs and the top HDC fired, then the particle also
must have crossed the lower HDC planes.
The result for the efficiency for one of these planes is shown in the
left part of Figure\,\ref{fig:efficiency},
the projection to the total efficiency of the 4-plane HDC package in
the right part.

HDC events are classified according to their hit pattern.
In the data analysis so-called $ok$  numbers are established.
Single hits and adjacent double hits get $1 \leq ok \leq 6$,
whereas $ok > 6$  is related to the occurence of non-adjacent double or 
multiple hits, electronic crosstalk, negative drift times, etc,
which go along with an increased error probability for the 
calculated trajectory.
The efficiency rises monotonically with the applied voltage if events with
any hit pattern ($ok \geq 1$) are taken into account.
This, however, is not the case for
the useful events with $1 \leq ok \leq 6$, where the efficiency shows a clear 
maximum around $3.1$\,kV.
This is due to the fact that multiple hits and crosstalk are much enhanced 
above the optimum high voltage.
The results depicted in Figure\,\ref{fig:efficiency} depend
on the particle ionization density and on the threshold of the 
amplifier/discriminator.
Furthermore, the efficiency depends on the orientation of the particle
trajectory relative to the HDC planes. 
This will be reconsidered as a source of false systematic asymmetries 
in the polarization measurements.

\section{Measurement of proton polarization}
\label{sec:polarization}

The detector setup in Spectrometer A (compare Figure\,\ref{fig:setup})
enables a measurement of the proton trajectories before and after scattering 
in the carbon analyzer.
Therefore the polar and azimuthal scattering angles can be determined as
required for the polarization analysis.
It is also possible to extract the position of the scattering vertex.
This is necessary for a separation of events scattered in the carbon
analyzer from those scattered e.g. in the scintillator planes of the 
spectrometer.
Furthermore, a diagnosis of errors both in the VDCs and in the HDCs
becomes possible \cite{Pospischil00}.

\begin{figure}
\centerline{\psfig{figure=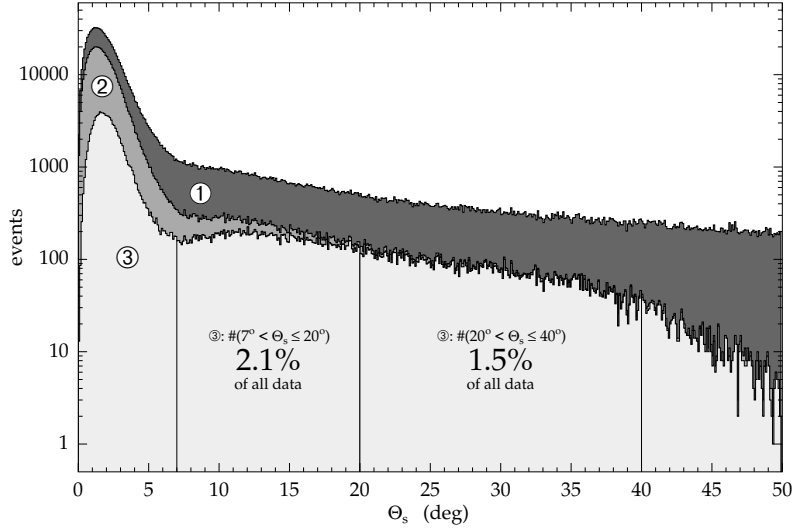,width=11.5cm}}
\caption{ Proton-carbon scattering angles for an analyzer thickness
          of 7\,cm and a proton energy range of 170 -- 260\,MeV.
          Curve 1 represents all data whereas in curve 2 events with drift 
          chamber errors are discarded. 
          In spectrum 3 a scattering vertex 
          within the carbon analyzer is required, which also suppresses
          events with small scattering angles.   }
\label{fig:theta}
\end{figure}

Figure\,\ref{fig:theta} shows the distribution of the polar scattering 
angle $\Theta_s$  measured with a 7\,cm thick carbon analyzer.
The proton kinetic energies varied between 170 and 260\,MeV across 
the acceptance of the spectrometer.
In contrast to other focal-plane polarimeters 
no small-angle rejection \cite{Lourie91} was used. 
Therefore the spectrum 
is dominated by small scattering angles $\Theta_s < 7^{\circ}$.
The efficiency of the polarimeter is related to the small fraction of 
$2.1$\,\% of events in the angular range 
$7^{\circ} < \Theta_s \leq 20^{\circ}$,
where the analyzing power is large and well known 
\cite{Aprile-Giboni83,McNaughton85}.
The analyzing power at larger angles is reconsidered in subsection
\ref{sec:A_large-angle}.

\begin{figure}
\centerline{\psfig{figure=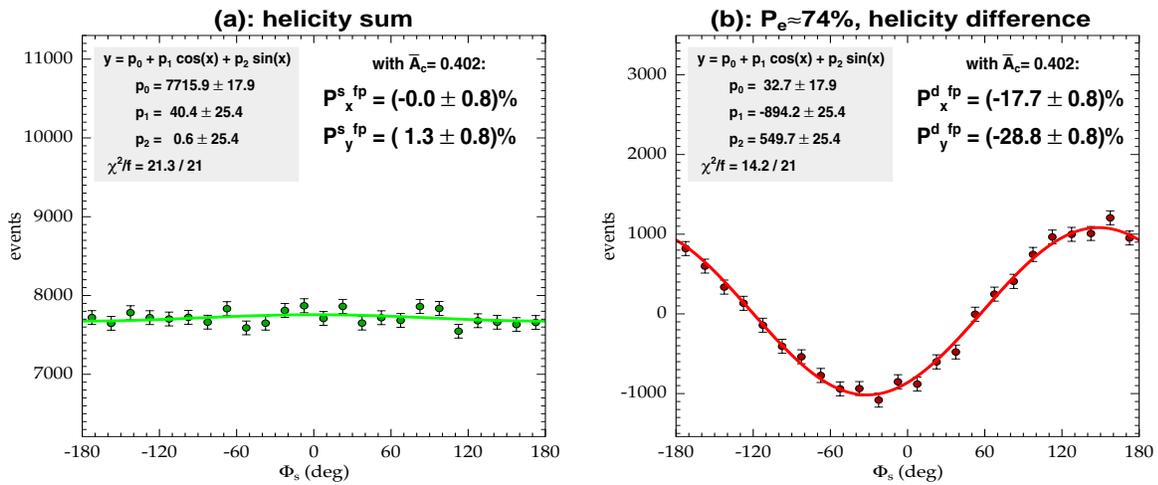,width=16cm}}
\caption{ Azimuthal angular distribution measured in the proton polarimeter
          for \peep elastic scattering. As explained in the text, the
          `helicity-sum' spectrum (a) is expected to be flat, whereas the
          `helicity-difference' (b) shows an asymmetry proportional to
          the recoil polarization. In this example the beam polarization 
          was $P_e \simeq 74$\,\%.}
\label{fig:phi_asym}
\end{figure}

The proton polarization is determined according to Eq.\,\ref{eq:sigma}
from the azimuthal angular distribution.
Such distributions are shown in Figure\,\ref{fig:phi_asym} from the 
\peep elastic scattering reaction for two cases labelled
`helicity-sum' (a) and `helicity-difference' (b).

In elastic electron-proton scattering the recoil proton polarization 
is proportional to the longitudinal polarization of the electron beam 
\cite{AR74,ACG81}
and thus flips sign under reversal of beam-helicity.
During the experiment the electron-beam helicity is flipped at the source
\cite{Aulenbacher97} on a random basis with a frequency of 1\,Hz.
Therefore the sum of events with positive and negative helicity corresponds to 
unpolarized beam and no azimuthal modulation must occur in 
Figure\,\ref{fig:phi_asym}(a).
In contrast, in the difference of the $\Phi$-distributions for positive and
negative beam helicities (Figure\,\ref{fig:phi_asym}(b)) the
asymmetries add up.
In the `helicity-difference' distribution instrumental asymmetries --
which of course are independent of beam helicity -- cancel out.
This is not the case for the `helicity-sum'.
Although there are no large instrumental effects visible in 
Figure\,\ref{fig:phi_asym}(a) the false systematic asymmetries are
analyzed in more detail in section\,\ref{sec:false}.
Eventwise calculation of the analyzing power according to the parameterization
\cite{McNaughton85} yielded for the data of Figure\,\ref{fig:phi_asym} 
a mean value of $\bar A_C = 0.402$.

In general, from the `helicity-sum' distribution it is possible to extract 
two beam-helicity {\em in}dependent recoil polarization components, 
while from the `helicity-difference' two beam-helicity dependent components
are obtained.
However, these polarizations are measured behind the spectrometer's 
focal plane, i.e. after spin precession in a magnetic system.
In order to determine the proton polarization at the electron scattering
vertex relative to the frame of the electron scattering plane, 
which is defined through the incident and scattered electron
momenta $\vec p_i$  and $\vec p_f$, respectively,
\begin{equation}
\hat z = \frac{\vec p_i - \vec p_f}{|\vec p_i - \vec p_f|}, \quad
\hat x = \frac{\vec p_i \times \vec p_f}{\vec p_i \times \vec p_f}, \quad
\hat y = \hat z \times \hat x,
\label{eq:sp-frame}
\end{equation}
the polarization measured in the focal plane must be traced back
through the fields of the spectrometer.
Despite the obvious complication, it is only through the spin precession
that the longitudinal polarization component 
(in the direction of the proton momentum at the electron vertex)
becomes accessible.

\subsection{Spin precession}

The description of the precession of a spin vector $\vec S$
in Spectrometer A is based on the Thomas equation \cite{Thomas27}.
For pure magnetic fields it can be cast into the form
\begin{equation}
\frac{d\vec S}{dt} = \frac{e}{m\gamma} \vec S \times
  \left[ \frac{g}{2} \vec B_{\parallel} + (1+\frac{g-2}{2}\gamma) \vec B_{\perp} \right],
\end{equation}
where $e$  and $m$  are the particle's charge and mass, and $g$  is its 
$g$-factor; $\gamma$  is the Lorentz-factor and the magnetic field is
split into two parts, $\vec B_{\parallel}$  and $\vec B_{\perp}$, 
which are parallel and perpendicular to the particle's momentum, respectively.

In its vertical midplane the QSDD-type Spectrometer A \cite{Blomqvist98}
can be approximated as a pure dipole with no longitudinal fields.
In this case for a Dirac particle with $g = 2$  the precession of the 
spin-vector is the same as for the momentum vector.
However, due to the large anomalous magnetic moment of the proton
($g = 5.586$) its spin precesses against its direction.
For a momentum of $p_p = 630$\,MeV/c the spin precession angles vary 
between $191^{\circ}$  and $242^{\circ}$  across the dispersive
(i.e. the vertical) acceptance of the spectrometer.
It is important to have the spin precession 
around $45^\circ$  (modulo $90^\circ$) in order to
achieve enough sensitivity to the polarization components both in
longitudinal and in dispersive direction at the electron vertex.

\begin{figure}
\centerline{\psfig{figure=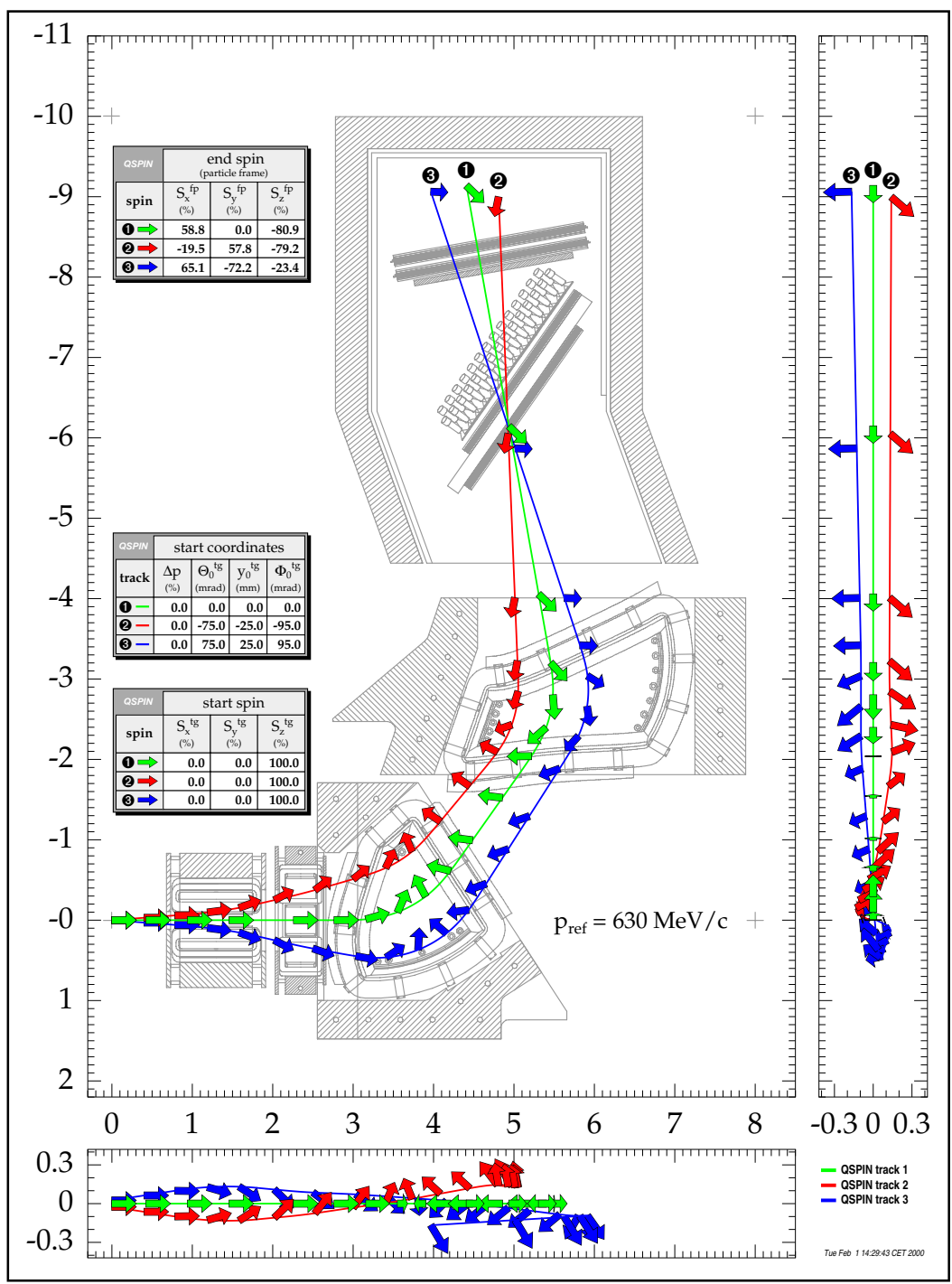,width=14cm}}
\caption{ Side view of the spin precession in Spectrometer A, calculated
          with \texttt{QSPIN} for three proton trajectories.
          Different spectrometer-target coordinates result in a completely
          different precession of the spins, 
          which initially were oriented in the same direction.
          Also shown are the rear and bottom projections of the proton 
          tracks and spins. The scale is in meters. }
\label{fig:spin_precession}
\end{figure}

In general, the spin precession through the spectrometer is complicated
due to the longitudinal fields and the varying bending directions
in the consecutive optical elements.
It is computed with the C++ code \texttt{QSPIN} \cite{Pospischil00}
which evolves both momentum and spin along the trajectories 
using a Runge-Kutta method with adaptive stepsize according to Cash and Karp
\cite{CK90,Pre95}.
\texttt{QSPIN} calculates the required magnetic field components 
similar to the \texttt{RAYTRACE} code \cite{Kowalski87},
which originally was used for the design of the spectrometer's optics.
The \texttt{QSPIN} results for protons of $p_p = 630$\,MeV/c with
spins oriented in $\hat z$  direction at the target are visualized in 
Figure\,\ref{fig:spin_precession} for three trajectories with different
so-called spectrometer-target coordinates $\Theta_0^{tg}$  (dispersive
angle), $y_0^{tg}$  (long target coordinate) and $\Phi_0^{tg}$  
(non-dispersive angle). 
Obviously, the different trajectories result in completely different
spin orientations behind the magnetic system.

Since the spin precession is a rotation of the initial spin
in the spectrometer-target frame (tg) into the final direction in the
particle frame behind the focal plane 
(fp, c.f. Figure\,\ref{fig:spin_precession}),
it can be written in matrix form:
\begin{equation}
\left( \begin{array}{c}   
S_x \\ S_y \\ S_z \\ 
\end{array} \right)^{fp} = 
\left( \begin{array}{ccc} 
M_{xx} & M_{xy} & M_{xz} \\ 
M_{yx} & M_{yy} & M_{yz} \\ 
M_{zx} & M_{zy} & M_{zz} \\ 
\end{array} \right) 
\left( \begin{array}{c}   
S_x \\ S_y \\ S_z \\ 
\end{array} \right)^{tg} \, . 
\label{eq:spintrans}
\end{equation}
Similar to the case of its optics matrix, the elements of the spectrometer's
spin transfer matrix (STM), $M_{\kappa\lambda}$,
can be expressed as polynomials in the spectrometer-target coordinates 
$\Theta_0^{tg}$, $y_0^{tg}$, $\Phi_0^{tg}$,  
the reference momentum setting $p_{ref}$  of the spectrometer and
the deviation $\Delta p$  of the particle's momentum from $p_{ref}$:
\begin{equation}
M_{\kappa\lambda} = \sum_{ijklm} 
\langle M_{\kappa\lambda} | \Delta p^i\,\Theta _0^j\,y_0^k\,\Phi _0^l p_{ref}^m\rangle
\, \Delta p^i\,\Theta _0^j\,y_0^k\,\Phi _0^l\,p_{ref}^m    
\label{eq:matrel}
\end{equation}
with $i, j, k, l, m \in N$ and $\kappa,\lambda \in {x,y,z}$.
The polynomial coefficients were determined 
by $\chi^2$  minimization of pseudo data.
Those were generated by \texttt{QSPIN} on 1715 different trajectories 
across the spectrometer's acceptance for each of three initial spin
orientations at the target and six different momentum settings
within $p_{ref} = 480$--$630$\,MeV/c 
(which covers the momentum range in which the polarimeter can be operated).

In a real experiment the polarization is measured behind the magnetic 
deflection and must then be traced back through the spectrometer.
However, the STM cannot be inverted directly, 
because only two polarization components are measurable in the polarimeter.
Nevertheless, there is a twofold redundancy that can be exploited:
\begin{enumerate}
\item{The electron-helicity dependent and independent 
      parts of the recoil polarization 
      can be separated by differences and sums 
      of the measured asymmetries, respectively 
      (compare beginning of section\,\ref{sec:polarization}).
      Symmetric averaging around the direction of momentum transfer
      yields for certain
      reactions (see for example \cite{HS98}) the two
      recoil polarization components in the electron scattering 
      plane (longitudinal and transversal), which are beam-helicity dependent,
      and the normal component, which is helicity independent. }
\item{Events from the same physical situation 
      (and thus with the same recoil polarization) 
      are obtained with different spectrometer-target coordinates due to, 
      e.g., the possible tilting of the electron scattering plane against
      Spectrometer A or the distribution of scattering vertices over
      the target length.
      The related large variation of the spin precession is exploited
      in the following fitting procedure. }
\end{enumerate}
The two focal plane polarization components are related to
the polarization at the scattering vertex 
(in the frame of Eq.\,\ref{eq:sp-frame})
by
\begin{equation}
\left(\begin{array}{c}
P_{x} \\
P_{y} \\
\end{array}\right)^{fp} =
\left(\begin{array}{ccc}
M_{xx} & M_{xy} & M_{xz} \\
M_{yx} & M_{yy} & M_{yz} \\
\end{array}\right) \cdot T_{\lambda \xi} \cdot
\left(\begin{array}{c}
P_{x} \\
P_{y} \\
P_{z} \\
\end{array}\right)^{sp},
\label{eq:sp_fp_transformation}
\end{equation}
or
\begin{equation}
P_{\kappa}^{fp} = \sum\limits_{\lambda = x}^{z} \sum\limits_{\xi = x}^{z}
M_{\kappa\lambda}(\Delta p, \Theta_{0}^{tg}, 
y_{0}^{tg}, \Phi_{0}^{tg}, p_{ref})\;
T_{\lambda\xi}(\Theta^{h}_{e}, \Phi^{h}_{e}, 
|\vec{p}_{f}|, |\vec{p}_{i}|, \Phi_{A})\;
P_{\xi}^{sp}\;,
\label{eq:sumsp2fpp}
\end{equation}
with $\kappa = x,y$. 
$T_{\lambda\xi} \,=\, \hat{\lambda}^{tg}\cdot\hat{\xi}^{sp}$  with
$\lambda,\xi \in \{x, y, z\}$  is the $3 \times 3$  matrix describing the
rotation between the coordinate frames of spectrometer-target and 
scattering plane.
The angles $\Theta^{h}_{e}$  and $\Phi^{h}_{e}$  characterize the direction
of the scattered electron and $\Phi_{A}$  is the central angle of 
Spectrometer A relative to the incident beam.
The matrix product yields the complete $3 \times 2$  imaging matrix 
$F_{\kappa\xi}(\vec{x}) = 
\sum\limits_{\lambda = x}^{z} M_{\kappa\lambda} \cdot T_{\lambda\xi}$
with
$\vec{x} = (\Delta p, \Theta_{0}^{tg}, y_{0}^{tg}, \Phi_{0}^{tg}, p_{ref},
\Theta^{h}_{e}, \Phi^{h}_{e}, |\vec{p}_{f}|, |\vec{p}_{i}|, \Phi_{A})$.

In order to enable fitting,
the acceptance in $\vec x$  is subdivided into $N_D$  bins 
for which the focal plane polarization is measured separately as
$P^{fp}_{x,i}(\vec{x_{i}})$ and $P^{fp}_{y,i}(\vec{x_{i}})$
with errors 
$\Delta P^{fp}_{x,i}(\vec{x_{i}})$ and $\Delta P^{fp}_{y,i}(\vec{x_{i}})$,
$i = 1,...,N_D$.
Under the assumption that the components
$P_{x,y,z}^{sp}$  themselves are independent of $\vec x$  
they can be determined by minimizing
\begin{equation}
\chi^{2} = \sum\limits_{i=1}^{N_{D}}\left(
\left[ \frac{P^{fp}_{x,i}(\vec{x_{i}}) - 
\sum\limits_{\xi = x}^{z} F_{x\xi}(\vec{x}_{i}) P_{\xi}^{sp}}
{\Delta P^{fp}_{x,i}(\vec{x_{i}})} \right]^{2} +
\left[ \frac{P^{fp}_{y,i}(\vec{x_{i}}) - 
\sum\limits_{\xi = x}^{z} F_{y\xi}(\vec{x}_{i})) P_{\xi}^{sp}}
{\Delta P^{fp}_{y,i}(\vec{x_{i}})} \right]^{2} \right) \, .
\label{eq:STMchisq}
\end{equation}
The requirements
$\frac{\partial \chi^2}{\partial P_{\xi}^{sp}} = 0$
lead to the matrix equation
\begin{equation}
A \cdot \vec{P}^{sp} = \vec{b},
\label{eq:STMAxb}
\end{equation}
where the elements of the $3\times 3$  matrix $A$ 
and of the vector $\vec{b}$ are given by
\begin{equation}
A_{\xi\mu} = \sum\limits_{i=1}^{N_{D}} \left(
\frac{F_{x\xi}(\vec{x}_{i})F_{x\mu}(\vec{x}_{i})}
{(\Delta P^{fp}_{x,i}(\vec{x_{i}}))^{2}} +
\frac{F_{y\xi}(\vec{x}_{i})F_{y\mu}(\vec{x}_{i})}
{(\Delta P^{fp}_{y,i}(\vec{x_{i}}))^{2}} \right),
\label{eq:STMA}
\end{equation}
\begin{equation}
b_{\mu} =
\sum\limits_{i=1}^{N_{D}} \left(
\frac{P^{fp}_{x,i}(\vec{x_{i}})F_{x\mu}(\vec{x}_{i})}
{(\Delta P^{fp}_{x,i}(\vec{x_{i}}))^{2}} +
\frac{P^{fp}_{y,i}(\vec{x_{i}})F_{y\mu}(\vec{x}_{i})}
{(\Delta P^{fp}_{y,i}(\vec{x_{i}}))^{2}} \right)\; .
\label{eq:STMchisqmin}
\end{equation}
The polarization is finally found as
\begin{equation}
\vec{P}^{sp} = A^{-1} \cdot \vec{b}
\label{eq:STMxAb}
\end{equation}
with the (correlated) error
\begin{equation}
\Delta P^{sp}_{\xi} = \sqrt{(A^{-1})_{\xi\xi}}.
\label{eq:errSTMxAb}
\end{equation}

\subsection{Elastic \peep measurements}
\label{sec:elastics}

The calculation of the spin precession with \texttt{QSPIN} and the trace back
of the polarization measured in the focal plane polarimeter 
to the electron vertex
was checked with elastic \peep measurements.
For a given degree of longitudinal electron polarization $P_e$, 
the recoil proton polarization is determined  \cite{AR74,ACG81} 
by electron kinematics and by
the proton's Sachs form factors $G_E$  and $G_M$, which, at low $Q^2$, are
known at the one-percent level:
\begin{eqnarray}
P_x^{sp} &=& P_e \frac{a G_E G_M}{G_E^2 + c G_M^2}, \label{eq:P_x_elast} \\
P_y^{sp} &=& 0,                                     \label{eq:P_y_elast} \\
P_z^{sp} &=& P_e \frac{b G_M^2}{G_E^2 + c G_M^2}.   \label{eq:P_z_elast} 
\end{eqnarray}
The axes are defined according to Eq.\,\ref{eq:sp-frame} and
the kinematical factors 
\begin{eqnarray}
a &=& -2\sqrt{\tau(1+\tau)} \tan \frac{\theta_e}{2}, \\
b &=& -a \sqrt{\tau(1+\tau \sin^2 \frac{\theta_e}{2})} / \cos\frac{\theta_e}{2}
      \qquad \mbox{and} \\
c &=& \tau + \frac{1}{2}a^2
\end{eqnarray}
are fixed by the electron scattering angle, $\theta_e$,  and the squared 
four-momentum transfer in units of the proton rest mass, $\tau = Q^2/4m_p^2$.
\begin{figure}[t]
\centerline{\psfig{figure=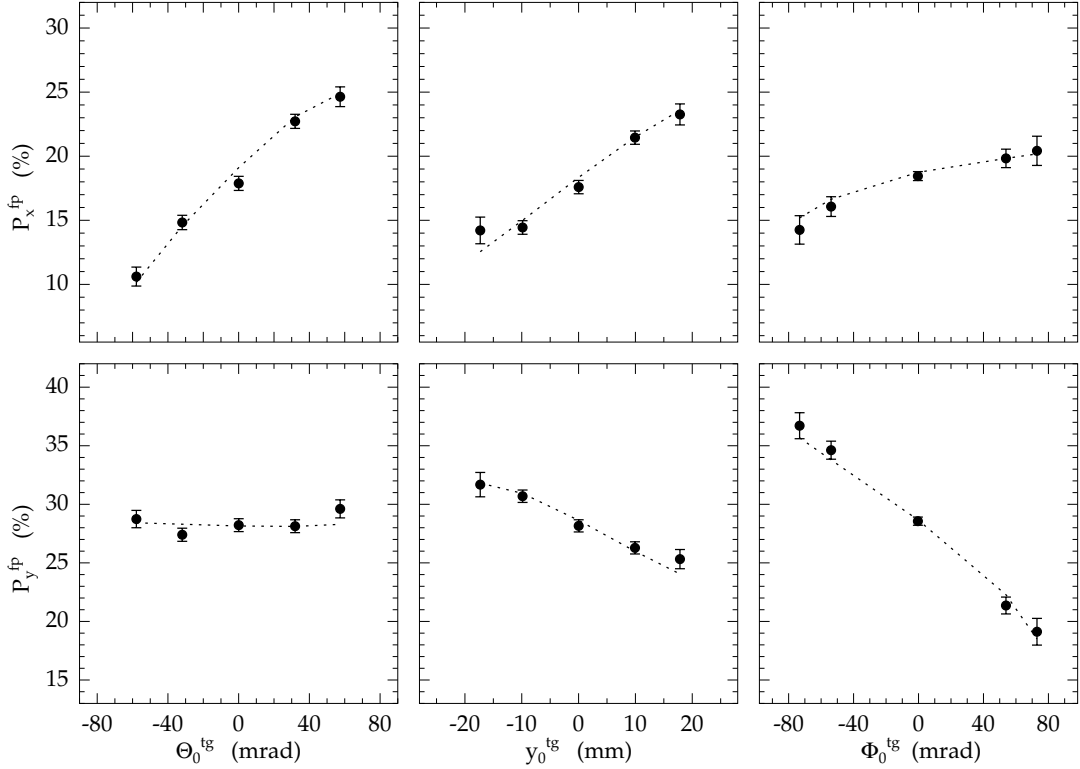,width=15cm}} 
\caption{ The two transverse polarization components after the
         spin precession in the spectrometer as a function of the measured
         dispersive angle (left), the target-length
         coordinate (middle) and the non-dispersive angle (right).
         The curves represent the \texttt{QSPIN} calculations 
         described in the text and the data points are the results of 
         the elastic $p(\vec e, e'\vec p\,)$  measurements. }
\label{fig:elastics}
\end{figure}

For the two transverse polarization components in the focal plane,
$P_x^{fp}$  and $P_y^{fp}$,
Figure\,\ref{fig:elastics} shows the comparison between 
\texttt{QSPIN} calculation (curves) and measurement (dots).
They agree very well as a function of the spectrometer-target coordinates
$\Theta_0^{tg}$, $y_0^{tg}$  and $\Phi_0^{tg}$.

From the polarization components in the scattering plane,
$P_x^{sp}$  and $P_z^{sp}$, it is possible,
as can be seen from Eqs.\,\ref{eq:P_x_elast} and \ref{eq:P_z_elast},
to determine the beam polarization independently of the 
proton form factors:
\begin{equation}
P_e = \frac{b}{a^2} \frac{(P_x^{sp})^2}{P_z^{sp}} + \frac{c}{b} P_z^{sp}.
\label{eq:P_e}
\end{equation}
The beam polarizations extracted from the measured proton polarization
components were confirmed within a relative error 
of $2.5$\,\% \cite{Groezinger00} by the new M{\o}ller polarimeter 
of the A1 collaboration at MAMI \cite{Bartsch00}.
This also confirms the absolute height of the proton-carbon analyzing power.

\subsubsection{Analyzing power at large scattering angles}
\label{sec:A_large-angle}

\begin{figure}[t]
\centerline{\psfig{figure=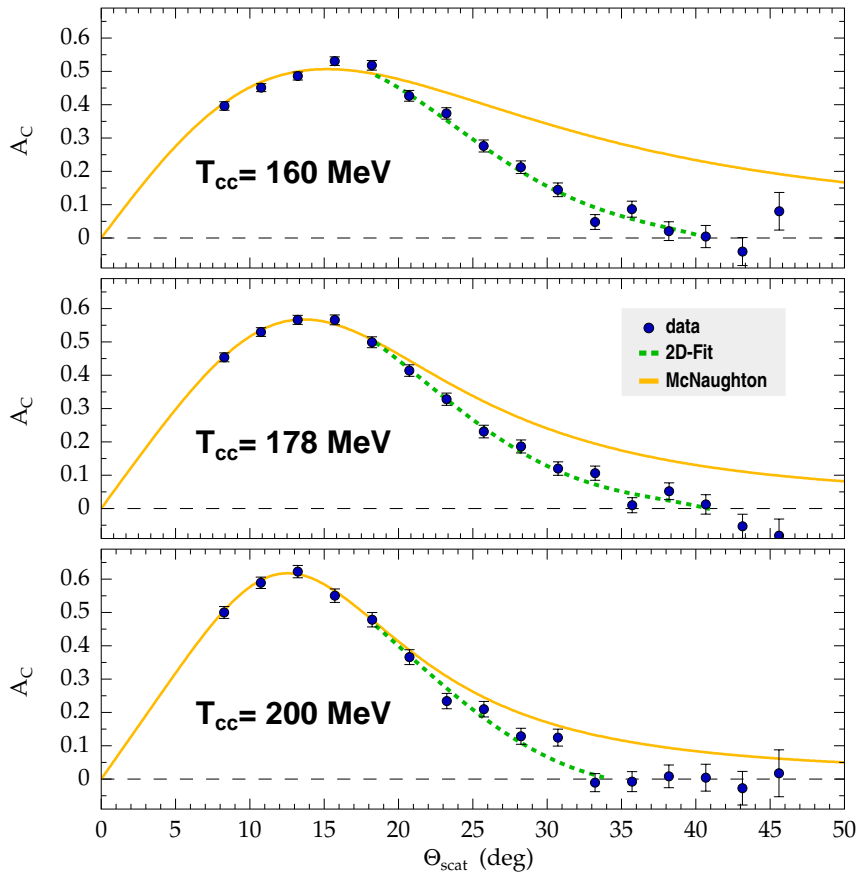,width=12cm}}
\caption{ Data of Table\,\ref{tab:A_C} (full circles) in comparison with the
          McNaughton parametrization \protect\cite{McNaughton85} (full curve).
          The broken curve represents the fit (Eq.\,\ref{eq:A_C-fit})
          to the data points. }
\label{fig:A_C}
\end{figure}

While in the relevant energy range below 250\,MeV the inclusive 
proton-carbon analyzing power is known at the 2\,\% level for scattering
angles up to 20$^{\circ}$, the accuracy is much lower for larger angles
\cite{Aprile-Giboni83,McNaughton85,Waters78}.
However, as can be seen from Figure\,\ref{fig:theta}, almost half of the
large-angle scattering events are in the range 20$^{\circ}$  -- 50$^{\circ}$
due to the large angular acceptance of the HDCs.
These data were used to determine the 
analyzing power for large scattering angles relative to the lower 
angular range.
The results are summarized in Table\,\ref{tab:A_C} 
and in Figure\,\ref{fig:A_C}
for three proton kinetic
energies in the center of the carbon analyzer, $T_{CC}$. 
The acceptance in $T_{CC}$  is approximately $\pm 10$\,MeV around 
the mean values.
\begin{table}
\begin{center}
\begin{tabular}{|r|r|r|r|}
\hline
 &
\multicolumn{1}{c|}{\footnotesize $T_{\rm CC}$~=~160 MeV} &
\multicolumn{1}{c|}{\footnotesize $T_{\rm CC}$~=~178 MeV} &
\multicolumn{1}{c|}{\footnotesize $T_{\rm CC}$~=~200 MeV} 
\\
\multicolumn{1}{|c|}{$\Theta_{\rm s}$} &
\multicolumn{1}{c|}{$A_{\rm C}$} &
\multicolumn{1}{c|}{$A_{\rm C}$} &
\multicolumn{1}{c|}{$A_{\rm C}$} \\
\hline
\hline
 8.3$^{\circ}$ &  0.396$~\pm$~ 0.013 &  0.454 ~$\pm$~ 0.013 &  0.500 ~$\pm$~ 0.018\\
10.7$^{\circ}$ &  0.451$~\pm$~ 0.012 &  0.530 ~$\pm$~ 0.013 &  0.589 ~$\pm$~ 0.017\\
13.2$^{\circ}$ &  0.486$~\pm$~ 0.012 &  0.566 ~$\pm$~ 0.014 &  0.623 ~$\pm$~ 0.018\\
15.7$^{\circ}$ &  0.531$~\pm$~ 0.013 &  0.566 ~$\pm$~ 0.015 &  0.550 ~$\pm$~ 0.020\\
18.2$^{\circ}$ &  0.518$~\pm$~ 0.015 &  0.499 ~$\pm$~ 0.016 &  0.478 ~$\pm$~ 0.022\\
20.7$^{\circ}$ &  0.427$~\pm$~ 0.016 &  0.414 ~$\pm$~ 0.017 &  0.366 ~$\pm$~ 0.022\\
23.2$^{\circ}$ &  0.374$~\pm$~ 0.017 &  0.328 ~$\pm$~ 0.018 &  0.234 ~$\pm$~ 0.023\\
25.7$^{\circ}$ &  0.276$~\pm$~ 0.018 &  0.231 ~$\pm$~ 0.019 &  0.210 ~$\pm$~ 0.023\\
28.2$^{\circ}$ &  0.212$~\pm$~ 0.019 &  0.186 ~$\pm$~ 0.020 &  0.128 ~$\pm$~ 0.024\\
30.7$^{\circ}$ &  0.145$~\pm$~ 0.021 &  0.120 ~$\pm$~ 0.020 &  0.124 ~$\pm$~ 0.025\\
33.2$^{\circ}$ &  0.048$~\pm$~ 0.022 &  0.106 ~$\pm$~ 0.021 & -0.011 ~$\pm$~ 0.027\\
35.7$^{\circ}$ &  0.087$~\pm$~ 0.024 &  0.010 ~$\pm$~ 0.023 & -0.008 ~$\pm$~ 0.030\\
38.2$^{\circ}$ &  0.021$~\pm$~ 0.028 &  0.052 ~$\pm$~ 0.025 &  0.008 ~$\pm$~ 0.034\\
40.7$^{\circ}$ &  0.004$~\pm$~ 0.033 &  0.012 ~$\pm$~ 0.029 &  0.004 ~$\pm$~ 0.040\\
43.1$^{\circ}$ & -0.041$~\pm$~ 0.042 & -0.053 ~$\pm$~ 0.036 & -0.027 ~$\pm$~ 0.050\\
45.6$^{\circ}$ &  0.080$~\pm$~ 0.056 & -0.081 ~$\pm$~ 0.050 &  0.017 ~$\pm$~ 0.070\\
\hline
\end{tabular}
\end{center}
\vspace*{2mm}
\caption{Inclusive p - $^{12}$C analyzing power as a function of scattering
         angle $\Theta_s$, measured with polarized protons from the 
         elastic \peep reaction. }
\label{tab:A_C}
\end{table}
Figure\,\ref{fig:A_C} shows that these data agree well with
the McNaughton parameterization \cite{McNaughton85} 
up to scattering angles of 20 degrees, the limit of its validity.
They are also in agreement with earlier large angle data \cite{Waters78}
with larger errors.
The result of the $\chi^2$  minimization of the two-dimensional polynomial
\begin{equation}
A_C(\Theta_s, A_{CC}) = \sum_{i=0}^2 \sum_{j=0}^4 
                        a_{ij} (T_{CC}/\mbox{MeV})^i (\Theta_s/\mbox{deg})^j
\label{eq:A_C-fit}
\end{equation}
for the data between $\Theta_s = 15.7$  and $43.1$  degrees is 
shown as broken line in Figure\,\ref{fig:A_C},
and the parameters $a_{ij}$  are given in
Table\,\ref{tab:A_C-fit}.
\begin{table}
\begin{center}
\begin{tabular}{|c||c|c|c|c|c|}
\hline
$a_{ij}$& $j=0$      & $j=1$       & $j=2$        & $j=3$        & $j=4$          \\
\hline
\hline
$i=0$  & -18.5902    & 1.47447     & -0.0184451   & -0.00084808  & 1.66656e-05  \\
$i=1$  & 0.162901    & -0.00897434 & -0.000186479 & 1.81209e-05  & -2.56371e-07 \\
$i=2$  & -0.00034948 & 1.18913e-05 & 1.10527e-06  & -6.23088e-08 & 8.07095e-10  \\
\hline
\end{tabular}
\end{center}
\vspace*{2mm}
\caption{Coefficients of the polynomial Eq.\,\ref{eq:A_C-fit} in the range
         $\Theta_s = 15.7^\circ - 43.1^\circ$  and 
         $T_{CC} = 160 - 200$\,MeV}
\label{tab:A_C-fit}
\end{table}

\subsubsection{False asymmetries}
\label{sec:false}

In order to avoid false asymmetries at the edges of the acceptance,
each event with scattering angles $\Theta_s$  and $\Phi_s$  is only
accepted, if in opposite azimuthal direction, $\Phi_s + \pi$, 
it would have been accepted, too.
This geometrical acceptance test does, however, not avoid artificial
asymmetries which are due to systematic efficiency variations.
As was mentioned in section\,\ref{sec:efficiency} the detection efficiency 
of the HDCs depends on the orientation of the proton tracks relative to 
the HDC planes.
This potentially produces false asymmetries.
In contrast to the beam-helicity dependent polarization components which
are extracted from the `helicity-difference' 
(compare Figure\,\ref{fig:phi_asym}), the `helicity-sum' is fully sensitive 
to such effects.

\begin{figure}
\centerline{\psfig{figure=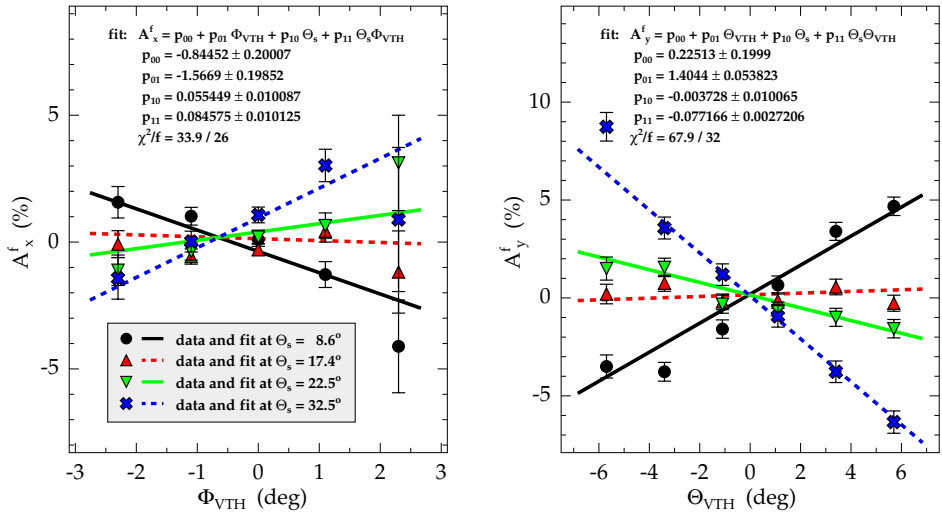,width=13cm}}
\caption{False asymmetries $A_x^f$  and $A_y^f$  as a function of the 
         orientation of the incoming proton tracks at the carbon analyzer,
         characterized through the angles $\Theta_{VTH}$  and $\Phi_{VTH}$.
         The straight lines represent linear fits.
         Different symbols and lines are used for scattering angles around
         $\Theta_s = 8.6^{\circ}, 17.4^{\circ}, 22.5^{\circ}$  and
         $32.5^{\circ}$  (fits include in addition data at 
         $\Theta_s = 12.4^{\circ}$  and $27.5^{\circ}$). }
\label{fig:false_asym}
\end{figure}

According to Eq.\,\ref{eq:P_y_elast} the beam-helicity independent 
recoil polarization must vanish in the elastic \peep reaction.
Therefore any measured `helicity-sum' asymmetry is false.
In Figure\,\ref{fig:false_asym} the false asymmetries $A_x^f$  and $A_y^f$  
are plotted as a function of the angles $\Theta_{VTH}$  and $\Phi_{VTH}$ 
which characterize the orientation of the incoming proton trajectory 
at the carbon analyzer.
The magnitude of the false asymmetries varies with the 
proton-carbon scattering angle $\Theta_s$.
It is described by two-dimensional linear fits,
the parameters of which are given as inserts in Figure\,\ref{fig:false_asym}.
$A_x^f(\Phi_{VTH},\Theta_s)$  and $A_y^f(\Theta_{VTH},\Theta_s)$  are used to 
correct $P_x^{fp}$  and $P_y^{fp}$, respectively.

\subsubsection{Systematic errors}

The false asymmetries discussed in the previous subsection only play a role in
the beam-helicity independent polarization components.
After correction, their remaining absolute contribution to the corresponding 
polarization components is less than 1\,\%.
The error in the analyzing power contributes with $\pm 2$\,\% relative.

A major part of the systematic uncertainty comes from the trace back
of the polarization through the spectrometer.
The quality of the STM and the consistency of the method are confirmed
within approximately 1\,\% through the elastic measurements 
and through the agreement of the extracted
electron-beam polarization with the
M{\o}ller measurements \cite{Groezinger00}.
In addition, the spin-precession calculation, and thus the trace back,
is affected by errors in the
spectrometer-target coordinates as determined by Spectrometer A.

Finally, if the recoil polarization is transformed into the electron 
scattering plane, then also errors from the electron arm contribute.
For the elastic \peep reaction, Table\,\ref{tab:syst_errors} gives 
a compilation of all systematic error contributions for
the (helicity-dependent) polarization components 
in the electron scattering plane 
as well as for the extracted beam polarization.
\begin{table}
\begin{center}
\begin{tabular}{|rl|c|c|c|}
\hline
 & & $P^{sp}_{x}$ (\%) & $P^{sp}_{z}$ (\%) & $P_{e}$ (\%)   \\
 \multicolumn{2}{|c|}{measured value} & -28.5 & 28.3 & 73.0 \\
\hline
\hline
 \multicolumn{2}{|c|}{\raisebox{-0.4ex}[0.4ex]{individual syst. errors}} & $\sigma^{sys}_{P^{sp}_{x}}$
 & $\sigma^{sys}_{P^{sp}_{z}}$ & $\sigma^{sys}_{P_{e}}$ \\[1ex]
\hline
 $\Delta\Delta p$~= & $\pm$ 0.2 \% & \kle $\pm$ 0.01 \,\, & $\mp$ 0.02 & \kle $\pm$
 0.01 \,\, \\
 $\Delta\Theta_{0}^{tg}$~= & $\pm$ 2 mrad & \kle $\pm$ 0.01 \,\, & $\mp$ 0.46 &
 $\mp$ 0.16 \\
 $\Delta y_{0}^{tg}$~= & $\pm$ 1.5 mm & $\mp$ 0.40 & $\mp$ 0.80 & $\pm$ 0.65 \\
 $\Delta\Phi_{0}^{tg}$~= & $\pm$ 2 mrad & $\mp$ 0.24 & $\mp$ 0.18 & $\pm$ 0.54
 \\
 $\Delta p_{ref}$~= & $\pm$ 0.5 MeV/c & \kle $\pm$ 0.01 \,\, & $\mp$ 0.02 &
 $\mp$ 0.01 \\
\hline
 $\Delta\Theta_{e}^{h}$~= & $\pm$ 2 mrad & \kle $\pm$ 0.01 \,\, & $\mp$ 0.12 &
 $\mp$ 0.03 \\
 $\Delta\Phi_{e}^{h}$~= & $\pm$ 2 mrad & $\pm$ 0.01 & $\mp$ 0.01 &
 $\pm$ 0.05 \\
 $\Delta|\vec{p}_{e'}|$~= & $\pm$ 0.2 MeV/c & $\pm$ 0.01 & $\pm$ 0.01 &
 $\mp$ 0.02 \\
 $\Delta|\vec{p}_{e}|$~= & $\pm$ 0.2 MeV/c & $\mp$ 0.01 & $\mp$ 0.01 &
 $\pm$ 0.01 \\
 $\Delta\Phi_{\rm A}$~= & $\mp$ 1 mrad & $\mp$ 0.03 & $\pm$ 0.03 &
 $\pm$ 0.06 \\
\hline
 $\Delta A_{\rm C}$~= & $\pm$ 2 \% (rel.) & $\pm$ 0.57 & $\pm$ 0.56 &
 $\pm$ 1.46 \\
\hline
\hline
 \multicolumn{2}{|c|}{total syst. error} & $\pm$ 0.74 & $\pm$ 1.10 & $\pm$
 1.70 \\
\hline
\hline
 \multicolumn{2}{|c|}{statistical error} & $\pm$ 0.43 & $\pm$ 0.67 & $\pm$
 1.02 \\
\hline
\end{tabular}
\end{center}
\vspace*{1mm}
\caption{Compilation of systematic errors for the elastic 
         \peep measurement with $T_{CC}=160$\,MeV. }
\label{tab:syst_errors}
\end{table}

In $P_x^{sp}$  and $P_z^{sp}$
the error is dominated by that of the long-target coordinate
$\Delta y_0^{tg}$  and to a lesser extent by the errors in the dispersive and
non-dispersive angles $\Delta \Theta_0^{tg}$  and $\Delta \Phi_0^{tg}$, 
respectively.
Also important is the uncertainty $\Delta A_C$  of the analyzing power,
which dominates the error of the extracted beam polarization $P_e$.

\section{Summary}
\label{sec:Summary}

Interesting nucleon and nuclear structure effects have recently
become accessible in double-polarization, exclusive electron scattering
experiments.
These experiments require in addition to the longitudinally 
polarized electron beam either a polarized target or recoil polarimetry.
For $(\vec e,e'\vec p\,)$-type coincidence experiments a focal plane 
polarimeter has been added to Spectrometer A of the 3-spectrometer setup of
the A1 collaboration at MAMI.
The proton polarization is measured through inclusive
proton-carbon scattering.
To this end, the standard VDC detector system has been supplemented by
a graphite analyzer of variable thickness and
two double planes of horizontal drift chambers to determine the trajectory
of the scattered protons against the incoming proton tracks,
which are measured in the VDCs.

The HDCs cover proton-carbon scattering angles up to 
45$^{\circ}$  over the whole area of the analyzer. 
They are operated with a gas mixture of 20\,\% argon and 80\,\% ethane.
Integration of the drift-time distribution for a homogeneously illuminated
HDC yields the drift-time to drift-distance relation. 
The left-right ambiguity of the HDC is resolved through readout of
the charge signals induced on adjacent potential wires by the ion drift
from an avalanche.
A position resolution of 300\,$\mu$m is achieved corresponding to an
angular resolution of 2\,mrad.

The measured polarization must be traced back through the magnetic fields of
the spectrometer. 
All three polarization components at the target can be determined 
simultaneously due to the variation of the spin precession across the
acceptance of the spectrometer and the redundancy provided by flipping
the electron-beam helicity.
The calculation of the precession was checked through the elastic \peep
reaction where the polarization transfer is determined by electron
kinematics and the (well known) proton elastic form factors.
These data were also used to determine the
analyzing power for scattering angles between $20^\circ$  and $45^\circ$
relative to the well known angular range below 20$^{\circ}$.
The absolute calibration of the polarimeter was confirmed
by M{\o}ller measurements of the beam polarization.

\section{Acknowledgements}
~\\

We thank H. Euteneuer and K.H. Kaiser and their staff for the perfect operation
of the accelerator as well as K. Aulenbacher and his group for running
the polarized source. 
For the engagement of the Mainz workshops we vicariously thank
R. B{\"o}hm, G. Jung and K.H. Luzius.

This work was supported by the Deutsche Forschungsgemeinschaft
within the SFB 443, the Schweizerische Nationalfonds and the
U.S. National Science Foundation.

\end{document}